\documentclass[prd,onecolumn,nopacs,nofootinbib]{revtex4}

\usepackage{amsmath}
\usepackage{graphicx}
\usepackage{amsfonts}
\usepackage{latexsym}
\usepackage{bbold}
\usepackage{wasysym}
\usepackage{calligra}
\usepackage{float}
\usepackage{ulem}
\usepackage{inputenc}
\usepackage{xspace}
\usepackage{url}
\usepackage{epstopdf}
\usepackage{tikz}

\usepackage{amsthm}

\newtheorem{theorem}{Theorem}

\newcommand{\be}{\begin{equation}}
\newcommand{\ee}{\end{equation}}
\newcommand{\beq} {\begin{equation}}
\newcommand{\eeq} {\end{equation}}
\newcommand{\ba}{\begin{eqnarray}}
\newcommand{\ea}{\end{eqnarray}}

\begin{document}

	\title{Metric-Affine Cosmologies: Kinematics of Perfect (Ideal) Cosmological Hyperfluids and First Integrals}
	
	\author{Damianos Iosifidis}
	\affiliation{Laboratory of Theoretical Physics, Institute of Physics, University of Tartu, W. Ostwaldi 1, 50411 Tartu, Estonia.}
	\email{damianos.iosifidis@ut.ee}
	
	\date{\today}
	\begin{abstract}
		
		We consider a generic  Metric-Affine Cosmological setup and  classify some particularly interesting specific cases of Perfect Hyperfluids. In particular, we present the form of conservation laws for the cases of pure spin, pure dilation and pure shear fluids. We also develop the concept of an incompressible hyperfluid and pay special attention to the case of a hypermomentum preserving hyperfluid. We also give a specific example on the emergence of the spin, dilation and shear currents through matter-connection couplings. In addition, starting from the  generalized acceleration equation for the scale factor including torsion and non-metricity we provide a first integral of motion relating the latter with the rest of the hyperfluid variables. These results then formalize the analysis of the non-Riemannian effects in Cosmology. 
		
	\end{abstract}
	
	\maketitle
	
	\allowdisplaybreaks
	
	
	\tableofcontents
	
	\section{Introduction}
	\label{intro}
	Non-Riemannian effects (i.e. spacetime torsion and non-metricity) of spacetime naturally arise when one considers matter configurations with microstructure \cite{hehl1995metric,puetzfeld2008probing}. In particular such effects could have a strong impact in the early time evolution of the Universe where the density of matter  was enormous. Under such extreme conditions the microscopic characteristics of matter (i.e. spin, dilation and shear \cite{hehl1995metric})   become  dominant and affect in a non-trivial way the evolution of the Universe. Furthermore, these features can also be dominant today and leave their footprint in regions of the Universe where the particle's density is high.

The aforementioned effects of the microstructure are nicely encapsulated in the so-called hypermomentum current of matter \cite{hehl1976hypermomentum} which is one of the sources of Metric-Affine Gravity (MAG) \cite{hehl1999metric}\footnote{The MAG framework has attracted a lot of attention lately \cite{Iosifidis:2021bad,Iosifidis:2021fnq,Iosifidis:2021tvx,vitagliano2011dynamics,sotiriou2007metric,percacci2020new,Jimenez:2020dpn,BeltranJimenez:2019acz,aoki2019scalar,Cabral:2020fax,Yang:2021fjy,Ariwahjoedi:2021yth,Rigouzzo:2022yan,bahamonde2020new,Bahamonde:2022meb,Bahamonde:2022kwg,shimada2019metric,Yang:2021fjy,kubota2021cosmological,Kubota:2020ehu,Mikura:2020qhc,Mikura:2021ldx,Boudet:2022nub}}. Recall that in the MAG framework, the affine connection is treated as an independent object and the field equations are obtained by varying both with respect to the metric and the affine connection. The hypermomentum tensor is then formally defined as the variational derivative of the matter part of the action with respect to the affine connection and can be split into its three pieces of spin, dilation and shear \cite{hehl1999metric}. The conservation laws obeyed by the hypermomentum and the rest of the energy tensors of MAG, namely the canonical and the metrical energy-momentum tensors, are derived, as usual, from the diffeomorphism invariance of the matter action and also from the GL invariance of the latter when working with exterior forms. 

Now, since matter with hypermomentum is to be understood as an extension of classical matter with its microstructure included, a fairly reasonable question to ask is what would be the generalization of a Perfect Fluid continuum possessing also non-trivial hypermomentum. If we recall that a Perfect Fluid can be defined as the fluid for which there exists a certain velocity with respect to which the fluid looks isotropic\footnote{This definition is given in Weinberg's book \cite{weinberg1972gravitation}.}, then  the extension would now consist in formulating a fluid, that apart from the classical energy-momentum tensor,  its associated hypermomentum tensor will also respect spatial isotropy. Such a construction was recently developed in \cite{Iosifidis:2020gth} and was further generalized in \cite{Iosifidis:2021nra}. This generalization of the Perfect Fluid notion, having also microstructure,   whose associated energy tensors all respect spatial isotropy is called the  Perfect Hyperfluid. Then, combining the conservation laws of MAG with the energy tensors of the Perfect Hyperfluid, one arrives at the evolution equations governing the behaviour of the latter.

 In general, the conservation laws of the Perfect Cosmological Hyperfluid are rather involved containing mixings of the  of the usual Perfect Fluid characteristics (i.e. isotropic density and pressure) and the hypermomentum degrees of freedom (i.e. the fluid's microstructure, see \cite{Iosifidis:2020gth,Iosifidis:2021nra}). More precisely, the continuity equation governing the density evolution of the fluid, now receives modifications associated to the micro-properties of matter \cite{Iosifidis:2020gth}. In addition, the evolution equations for hypermomentum variables contain in them also the density and pressure of the fluid. They do however, become  completely free from the latter and evolve separately for the case of a hypermomentum preserving  Perfect Hyperfluid \cite{Iosifidis:2020gth}. As we shall show, for specific forms of hyperfluids, the aforementioned conservation laws acquire relatively simple forms and in some cases  give quite some information for the whole kinematics.  

The paper is organized as follows. Firstly, we setup the notation and conventions and briefly discuss some geometrical and physical characteristics of MAG. Subsequently, we focus on the homogeneous and isotropic forms of torsion and non-metricity and categorize  some particular subcases of them related to certain geometries. We then recall the basic features of Perfect Cosmological Hyperfluids and analyse separately the three cases of pure shear, pure spin and pure dilation Hyperfluids. In particular, the form of the associated conservation laws of the fluid are obtained for each case, and a further classification for the case of Hypermomentum preserving fluids is considered. We also define the concept of in-compressible Hyperfluids, and give an example of a Hyperfluid whose hypermomentum is derived from a scalar field action that also contains couplings with the affine connection. Finally, starting from the acceleration equation, we derive in full generality a first integral of motion restricting the Hyperfluid motion in homogeneous Cosmologies. We then recap our results and discuss their various applications.

	\section{The Setup}

	\subsection{Geometric Setup}
		Let us briefly introduce here some basic definitions and conventions we shall be using throughout. We adopt the ones that appear in \cite{iosifidis2019metric}.  We consider a general n-dimensional Non-Riemannian space, endowed with a metric $g=g_{\mu\nu}dx^{\mu}dx^{\nu}$, $\mu,\nu=0,1,2,...,n-1$ and a generic affine connection $\nabla$ with coefficients $\Gamma^{\lambda}_{\;\;\mu\nu}$. This connection will in general be neither symmetric nor metric compatible and consequently we define the torsion and non-metricity tensors according to
\beq
S_{\mu\nu}^{\;\;\;\;\lambda}:=\Gamma^{\lambda}_{\;\;\;[\mu\nu]}
\eeq
\beq
Q_{\alpha\mu\nu}=-\nabla_{\alpha}g_{\mu\nu}=-\partial_{\alpha}g_{\mu\nu}+\Gamma^{\lambda}_{\;\;\mu\alpha}g_{\lambda\nu}+\Gamma^{\lambda}_{\;\;\nu\alpha}g_{\lambda\mu}
\eeq
  respectively. Taking contractions,  from non-metricity we can construct two vectors
\beq
Q_{\mu}:=Q_{\mu\alpha\beta}g^{\alpha\beta}\;\;, \;\; q_{\mu}:=Q_{\alpha\nu\mu}g^{\alpha\nu}
\eeq
with the one at the left oftentimes referred to as the Weyl vector. As for torsion we can define a vector along with a pseudo-vector, as follows
\beq
S_{\mu}:=S_{\mu\alpha}^{\;\;\;\alpha}\;\;, \;\; t_{\mu}:=\epsilon_{\mu\nu\alpha\beta}S^{\nu\alpha\beta}
\eeq
  with the latter being defined only for $n=4$.

  Additionally, the tensor
	\begin{equation}
R^{\mu}_{\;\;\;\nu\alpha\beta}:= 2\partial_{[\alpha}\Gamma^{\mu}_{\;\;\;|\nu|\beta]}+2\Gamma^{\mu}_{\;\;\;\rho[\alpha}\Gamma^{\rho}_{\;\;\;|\nu|\beta]}
\end{equation}
is the curvature tensor of the connection. Here  horizontal bars around an index denote that this index is left out of the (anti)-symmetrization. In non-Riemannian Geometries the only symmetry the latter possesses is antisymmetry in its last two indices as follows immediately by the above definition.  Without the use of any metric, we can construct the  two independent contractions for the curvature tensor
\beq
R_{\nu\beta}:=R^{\mu}_{\;\;\;\nu\mu\beta}\;,\; \widehat{R}_{\alpha\beta}:=R^{\mu}_{\;\;\;\mu\alpha\beta}
\eeq
The former defines as usual the Ricci tensor while the latter is the so-called  homothetic curvature tensor and is of purely non-Riemannian origin\footnote{In particular, it can be easily shown to be the field strength of $\Gamma^{\lambda}_{\;\;\lambda\mu}$.}. Once a metric is given we can form yet another  contraction
\beq
\breve{R}^{\lambda}_{\;\;\kappa}:=R^{\lambda}_{\;\;\mu\nu\kappa}g^{\mu\nu}
\eeq
which is oftentimes referred to as the co-Ricci tensor. However, the Ricci scalar is still uniquely defined since
\beq
R:=R_{\mu\nu}g^{\mu\nu}=-\breve{R}_{\mu\nu}g^{\mu\nu}\;,\;\; \widehat{R}_{\mu\nu}g^{\mu\nu}=0
\eeq
Now, it is a trivial  to show that the general affine connection can be decomposed as (see for instance \cite{iosifidis2019metric}) 
\begin{equation}\label{decgamma}
{\Gamma^\lambda}_{\mu \nu} = \tilde{\Gamma}^\lambda_{\phantom{\lambda} \mu \nu} + {N^\lambda}_{\mu \nu}\,,
\end{equation}
where
\beq\label{lcconn}
\tilde{\Gamma}^\lambda_{\phantom{\lambda}\mu\nu} = \frac12 g^{\rho\lambda}\left(\partial_\mu 
g_{\nu\rho} + \partial_\nu g_{\rho\mu} - \partial_\rho g_{\mu\nu}\right)
\eeq
is the usual Levi-Civita connection and the tensor ${N^\lambda}_{\mu\nu}$ is given 
\beq\label{distortion}
{N^\lambda}_{\mu\nu} = {\frac12 g^{\rho\lambda}\left(Q_{\mu\nu\rho} + Q_{\nu\rho\mu}
- Q_{\rho\mu\nu}\right)} - {g^{\rho\lambda}\left(S_{\rho\mu\nu} +
S_{\rho\nu\mu} - S_{\mu\nu\rho}\right)} \,.
\eeq	
and is the so-called distrortion tensor quantifying the departure from the Riemannian geometry.

	\subsection{Isotropic Torsion, Non-metricity and Distortion}

	In a homogeneous and isotropic FLRW spacetime\footnote{The metric in this setting is of course the usual Robertson-Walker which in the flat case reads: $ds^{2}=-dt^{2}+a^{2}(t)\delta_{ij}dx^{i}dx^{j}$,  $a(t)$ being the scale factor.} torsion contributes two degrees \cite{tsamparlis1979cosmological} of freedom (in $4$-dimensions) while non-metricity contributes  three \cite{minkevich1998isotropic}. The covariant form of the torsion and non-metricity tensors  that respect spatial isotropy and homogeneity, then read \cite{Iosifidis:2020gth}
\beq
S_{\mu\nu\alpha}^{(n)}=2u_{[\mu}h_{\nu]\alpha}\Phi(t)+\epsilon_{\mu\nu\alpha\rho}u^{\rho}P(t)\delta_{n,4} \label{isotor}
\eeq
and 
	\beq
	Q_{\alpha \mu \nu}  = A(t) u_\alpha h_{\mu \nu} + B(t) h_{\alpha(\mu} u_{\nu)} + C(t) u_\alpha u_\mu u_\nu
	\eeq
 respectively and $\delta_{n,4}$ is the Kronecker tensor ensuring that the pseudo-vectorial degree of freedom is there only in four dimensions. Here $\Phi, P$ are the degrees of freedom of torsion
	while $A,B$ and $C$ are the three degrees of freedom fully describing the evolution of non-metricity in such a setting. In addition the distortion tensor now contains five degrees of freedom and its covariant form reads,
	\begin{gather}
	N_{\alpha\mu\nu}^{(n)}=X(t)u_{\alpha}h_{\mu\nu}+Y(t)u_{\mu}h_{\alpha\nu}+Z(t)u_{\nu}h_{\alpha\mu}
	+V(t)u_{\alpha}u_{\mu}u_{\nu} +\epsilon_{\alpha\mu\nu\lambda}u^{\lambda}W(t)\delta_{n,4}
	\end{gather}	
	where the functions $X(t),Y(t),Z(t),V(t),W(t)$ represent the five degrees of of freedom of the distortion.
 Then, using the relations
\beq
Q_{\nu\alpha\mu}=2 N_{(\alpha\mu)\nu} \;\;, \;\;
S_{\mu\nu\alpha}=N_{\alpha[\mu\nu]}
\eeq	
one can connect the torsion and non-metricity variables to the distortion degrees of freedom according to 
\beq
2(X+Y)=B \;, \;\; 2Z=A\;, \;\; 2V=C \;, \;\; 2\Phi =Y-Z\;, \;\; P = W	 \label{dv}
\eeq
with inverse form
\beq
W=P \;, \;\; V=C/2 \;, \;\; Z=A/2	
\eeq
\beq
Y=2\Phi +\frac{A}{2}	\;\;, \;\;\; X=\frac{B}{2}- 2 \Phi -\frac{A}{2}
\eeq
 Whichever set one uses is then totally irrelevant to the actual physics; the first set (i.e. torsion/non-metricity variables) has a more transparent geometrical interpretation while the latter (i.e. distortion) is oftentimes more convenient for actual calculations.

 Now, there are $3$ special types of non-metricity that are of particular interest, these are the $i)$ Weyl, $ii)$ Fixed Length Vector and $iii)$ Totally Symmetric Traceless non-metricity. For the above Cosmological form, these constrained geometries put certain relations among $A,B$ and $C$. In particular we have the following:
	
	\begin{enumerate}
	    \item \underline{Weyl Non-metricity}: The constrained non-metricity of this type has the form $Q_{\alpha\mu\nu}=\frac{Q_{\alpha}}{n}g_{\mu\nu}$, implying the relations
	     \beq
	   B=0 \;\;, \;\;  C=-A
	    \eeq
	    and as a result
	    \beq
	    Q_{\alpha\mu\nu}=A\Big( u_{\alpha}h_{\mu\nu}-u_{\alpha}u_{\mu}u_{\nu} \Big)
	    \eeq
	    We can also have an extended version where B is allowed to have a non-zero value, i.e.,
     Extended Weyl Non-metricity: $C=-A$, $B \neq 0$.

	    \item \underline{Fixed Length Vector Non-metricity}: In this case lengths of vectors are preserved under parallel transport and this property is expressed by the demand $Q_{\alpha\mu\nu}u^{\alpha}u^{\mu}u^{\nu}=0$ (see the classic \cite{schrodinger1985space}) , yielding
	    \beq
	    B=-A \;\;, \;\; C=0
	    \eeq
	    and non-metricity takes the form
	    \beq
	  Q_{\alpha\mu\nu} = A\Big(  u_\alpha h_{\mu \nu} - h_{\alpha(\mu} u_{\nu)} \Big)
	    \eeq
	    
	    \item \underline{Totally Symmetric Traceless Non-metricity}: In this case, demanding $Q_{\alpha}=0$ and $q_{\alpha}=0$ together imply
	    \beq
	B=2A \;\;,  \;\; C=(n-1)A    \label{totsym}
	    \eeq
	   and the totally symmetric traceless non-metricity reads
	   \beq
	   Q_{\alpha\mu\nu}=Q_{(\alpha\mu\nu)}=A\Big( u_{\alpha}h_{\mu\nu}+u_{\mu}h_{\alpha\nu}+u_{\nu}h_{\alpha\mu}+(n-1)u_{\alpha}u_{\mu}u_{\nu} \Big) \label{qtraceless}
	   \eeq
	    
	\end{enumerate}
	
	 As for torsion, apart from the trivial cases of a purely vectorial (i.e. $\Phi \neq 0$ and $P=0$) and purely pseudo-vectortial (i.e.  $\Phi =0$ and $P\neq 0$), interesting cases are those of self-dual and anti-self-dual configurations. In particular, we have the following.

	Given the fact that torsion is a $2$-form, in $4$ dimensions it is natural to ask for the existence of self-dual and anti-self-dual configurations. For a Euclidean signature we define the (anti)-self-dual torsion form as
	\beq
		\star T^{a}=\pm  T^{a}
	\eeq 
	where the minus stands for anti-self-dual and the plus for self-dual configurations. Here $T^{a}$ is the torsion $2$-form expressed as $T^{a}=-S_{\mu\nu}^{\;\;\; a} dx^{\mu}\wedge dx^{\nu}$\footnote{The minus sign here and the absence of the $1/2$ factor are purely conventional.} in local coordinates and $\star$ the Hodge star. For Minkowskian signatures our definition for (anti)-self-duality reads
	\beq
	\star T^{a}=\pm i T^{a}
	\eeq
	Using this last duality condition, for Minkowskian signatures, and the isotropic torsion expression (\ref{isotor}) we readily see that the (anti)-self-duality  imposes that
	\beq
	P=\pm i \Phi
	\eeq
	among the torsion functions, where the plus sign corresponds to self-duality and the minus to anti-self-duality respectively. The full torsion tensor then reads,
 
  \underline{Self-Dual/Anti-Self-Dual Torsion}:

\beq
S_{\mu\nu\alpha}^{(n)}=\Big( 2u_{[\mu}h_{\nu]\alpha}\pm i \epsilon_{\mu\nu\alpha} \delta_{n,4}\Big) \Phi
\eeq
where $\epsilon_{\mu\nu\alpha}=\epsilon_{\mu\nu\alpha\kappa}u^{\kappa}$.

Let us now turn our attention to the matter content of the Theory that sources the above forms of torsion and non-metricity.

	\section{Perfect (Ideal) Hyperfluids}
We shall briefly review here the concept of the Perfect Hyperfluid which was developed in \cite{Iosifidis:2020gth} and was further  generalized in \cite{Iosifidis:2021nra}. The latter represents a direct generalization of the usual Perfect Fluid construction with the microscopic characteristics of matter included, which are encoded in the hypermomentum tensor. The construction consists on demanding spatial isotropy for the energy related tensors and then combine these forms with the general Metric-Affine Conservation laws to find how the evolution of matter sources. 

Firstly, let us recall the sources of Metric-Affine Gravity. These are the  Canonical and Metrical Energy-Momentum Tensors, that are respectively given by\footnote{Here $e_{\mu}^{\;\;a}$ is as usual the vielbein which defines the metric through $g_{\mu\nu}=e_{\mu}^{\;\;a}e_{\nu}^{\;\;a}g_{ab}$ where $g_{ab}$ is the tangent space metric. If the vielbein is taken to be orthonormal, the latter becomes the Minkowski metric $\eta_{ab}$.}
\beq\label{cemt}
{t^\mu}_c := \frac{1}{\sqrt{-g}} \frac{\delta S_{\text{M}}}{\delta {e_\mu}^c} \,.
\eeq
and
\beq
T_{\mu \nu} := - \frac{2}{\sqrt{-g}} \frac{\delta S_{\text{M}}}{\delta g^{\mu \nu}} = - \frac{2}{\sqrt{-g}} \frac{\delta (\sqrt{-g} \mathcal{L}_{\text{M}})}{\delta g^{\mu \nu}} \,,
\eeq
 along with the Hypermomentum tensor  \cite{hehl1976hypermomentum,hehl1978hypermomentum}
 which is defined as the variation of the matter part of the action with respect to the connection, namely

\beq
{\Delta_\lambda}^{\mu \nu} := - \frac{2}{\sqrt{-g}} \frac{\delta S_{\text{M}}}{\delta {\Gamma^\lambda}_{\mu \nu}} = - \frac{2}{\sqrt{-g}} \frac{\delta (\sqrt{-g} \mathcal{L}_{\text{M}})}{\delta {\Gamma^\lambda}_{\mu \nu}} \,.
\eeq
The latter tensor is quite important and is ultimately related with the micro-properties of matter such as spin, dilation and shear. It can be split into these three irreducible pieces of spin, dilation and shear according to (see \cite{hehl1995metric})
\beq
\Delta_{\alpha\mu\nu}=\Sigma_{\alpha\mu\nu}+\frac{1}{n}g_{\alpha\mu}Q_{\nu}+\hat{\Delta}_{\alpha\mu\nu} \label{hypsplit}
\eeq
with
\beq
\sigma^{\mu\nu\alpha}:=\Delta^{[\mu\nu]\alpha} \;\;\; (Spin)
\eeq
\beq
\Delta^{\nu}:=\Delta^{\alpha\mu\nu}g_{\alpha\mu} \;\;\; (Dilation)
\eeq
\beq
\Sigma^{\mu\nu\alpha}:=\Delta^{(\mu\nu)\alpha}-\frac{1}{n}g^{\mu\nu}\Delta^{\alpha} \;\;\; (Shear)
\eeq

	Working in the exterior calculus framework, from the GL and Diffeomorphism invariance of the matter action one gets the conservation laws \cite{hehl1995metric,Iosifidis:2020gth,obukhov2013conservation} (expressed in a holonomic frame):

	\beq
	t^{\mu}_{\;\;\lambda}
	= T^{\mu}_{\;\;\lambda}-\frac{1}{2 \sqrt{-g}}(2S_{\nu}-\nabla_{\nu})(\sqrt{-g}\Delta_{\lambda}^{\;\;\mu\nu}) \label{cc1}
	\eeq
	\beq
	\frac{1}{\sqrt{-g}}(2S_{\mu}-\nabla_{\mu})(\sqrt{-g}t^{\mu}_{\;\;\alpha})=-\frac{1}{2} \Delta^{\lambda\mu\nu}R_{\lambda\mu\nu\alpha}+\frac{1}{2}Q_{\alpha\mu\nu}T^{\mu\nu}+2 S_{\alpha\mu\nu}t^{\mu\nu} \label{cc2}
	\eeq
This is the most general form of the conservation laws of Metric-Affine Gravity. If we demand isotropy, then $t_{\mu\nu}$  must also be symmetric. On this assumption and using the trivial  identity
	\beq
	-\frac{1}{\sqrt{-g}}(2S_{\mu}-\nabla_{\mu})(\sqrt{-g}C^{\mu}_{\;\;\nu})=\tilde{\nabla}_{\mu}C^{\mu}_{\;\; \nu}-\frac{1}{2}Q_{\nu\alpha\beta}C^{\alpha\beta}-2 S_{\nu\alpha\beta}C^{\alpha\beta}
	\eeq
holding true  for any symmetric rank-$2$ tensor $C_{\mu\nu}$, the conservation laws simplify to
	\beq
	\tilde{\nabla}_{\mu}t^{\mu}_{\;\; \alpha}=\frac{1}{2}\Delta^{\lambda\mu\nu}R_{\lambda\mu\nu\alpha}+\frac{1}{2}Q_{\alpha\mu\nu}(t^{\mu\nu}-T^{\mu\nu}) \label{CL1}
	\eeq
	\beq 
	t^{\mu}_{\;\;\lambda}
	= T^{\mu}_{\;\;\lambda}-\frac{1}{2 \sqrt{-g}}\hat{\nabla}_{\nu}(\sqrt{-g}\Delta_{\lambda}^{\;\;\mu\nu}) \label{CL2}
	\eeq
	 The associated canonical and metrical energy-momentum tensors, of the Perfect Hyperfluid read \cite{Iosifidis:2021nra}
\beq
t_{\mu\nu}=\rho_{c}u_{\mu}u_{\nu}+p_{c} h_{\mu\nu} \label{canonical}
\eeq
\beq
T_{\mu\nu}=\rho u_{\mu}u_{\nu}+p h_{\mu\nu} \label{metrical}
\eeq
where $\rho_{c},p_{c}$ are the density and pressure associated with the canonical part and $\rho,p$ the usual density and pressure variables associated to $T_{\mu\nu}$. In the special case for which
\beq
t_{\mu\nu}=T_{\mu\nu}
\eeq
 we consequently have 
 \beq
 (2S_{\nu}-\nabla_{\nu})(\sqrt{-g}\Delta_{\lambda}^{\;\;\mu\nu})=0
 \eeq
 as seen from (\ref{cc1}), and the fluid obeying this condition is called \underline{Hypermomentum Preserving}. As we shall show in what follows this condition is possible for pure shear or pure dilation Hyperfluids but it is incompatible for pure spin fluids. 

In addition, demanding only spatial isotropy  the hypermomentum tensor of the Perfect Hyperfluid takes the covariant form \cite{Iosifidis:2020gth}	
	\beq
	\Delta_{\alpha\mu\nu}^{(n)}=\phi h_{\mu\alpha}u_{\nu}+\chi h_{\nu\alpha}u_{\mu}+\psi u_{\alpha}h_{\mu\nu}+\omega u_{\alpha}u_{\mu} u_{\nu}+\delta_{n,4}\epsilon_{\alpha\mu\nu\kappa}u^{\kappa}\zeta \label{Dform}
	\eeq
where $\phi,\chi,\psi,\omega$ and $\zeta$ are the functions encoding the microscopic characteristics of the fluid. These, along with $\rho,p,\rho_{c}$ and $p_{c}$ are in general spacetime functions. The set of equations (\ref{CL1})-(\ref{Dform}) exactly describes the behaviour of the Perfect Hyperfluid,  providing a direct generalization of the Perfect Fluid continuum when the intrinsic characteristics of matter (i.e. $\Delta_{\alpha\mu\nu}$) are also taken into account.  
In the above, as usual, $u^{\mu}$ is the normalized $n-velocity$  field ($u_{\mu}u^{\mu}=-1$) and we have performed an $1+(n-1)$ spacetime split with the  introduction of the projection tensor $h_{\mu\nu}=g_{\mu\nu}+u_{\mu}u_{\nu}$. For the above form of isotropic hypermomentum, its three irreducible parts of spin dilation and shear are respectively given by
\beq
\Delta_{[\alpha\mu]\nu}=(\psi-\chi)u_{[\alpha}h_{\mu]\nu}+\delta_{n,4}\epsilon_{\alpha\mu\nu\kappa}u^{\kappa}\zeta \label{spin}
\eeq
\beq
\Delta_{\nu}:=\Delta_{\alpha\mu\nu}g^{\alpha\mu}=\Big[ (n-1) \phi -\omega\Big] u_{\nu} \label{dil}
\eeq
\beq
\breve{\Delta}_{\alpha\mu\nu}=\Delta_{(\alpha\mu)\nu}-\frac{1}{n}g_{\alpha\mu}\Delta_{\nu} =\frac{(\phi+\omega)}{n}\Big[ h_{\alpha\mu}+(n-1)u_{\alpha}u_{\mu} \Big] u_{\nu} +(\psi +\chi)u_{(\mu}h_{\alpha)\nu} \label{shear}
\eeq

	 If we now impose homogeneity as well, that is if we consider an FLRW Universe, all  variables of the hyperfluid will be time dependent only and in this case our conservation laws (\ref{CL1}) and (\ref{CL2}) boil down to\footnote{As usual, the dot denotes time derivative.}
	\beq
	\Big[ \dot{\rho}+(n-1)H(\rho+p) \Big] u_{\nu}+(\rho +p)u^{\mu}\widetilde{\nabla}_{\mu}u_{\nu}=\frac{1}{2}u^{\mu}(\phi \widehat{R}_{\mu\nu}+\chi R_{\mu\nu}+\psi \breve{R}_{\mu\nu})+\frac{1}{2}u_{\nu}\Big[ (\rho_{c}-\rho)C+(p_{c}-p)(n-1)A \Big] \label{cl1}
	\eeq
	\begin{gather}
	-\delta^{\mu}_{\lambda}    \frac{\partial_{\nu}(\sqrt{-g}\phi u^{\nu})}{\sqrt{-g}}-u^{\mu}u_{\lambda}      \frac{\partial_{\nu}\Big(\sqrt{-g}(\phi+\chi +\psi +\omega) u^{\nu}\Big)}{\sqrt{-g}}
	\nonumber \\
	+\left[ \Big(2 S_{\lambda}+\frac{Q_{\lambda}}{2}\Big)u^{\mu}-\nabla_{\lambda}u^{\mu} \right]\chi +\left[ \Big(2 S^{\mu}+\frac{Q^{\mu}}{2}-\tilde{Q}^{\mu}\Big)u_{\lambda}-g^{\mu\nu}\nabla_{\nu}u_{\lambda}\right]\psi
	\nonumber \\
	+ u^{\mu}u_{\lambda}(\dot{\chi}+\dot{\psi}) -(\phi+\chi+\psi+\omega)(\dot{u}^{\mu}u_{\lambda}+u^{\mu}\dot{u}_{\lambda}) 
	=2(\rho -\rho_{c})u_{\lambda}u^{\mu}+2 (p-p_{c})h_{\lambda}^{\;\;\mu}   \label{conl2}
	\end{gather}
	Contracting the latter with $u^{\nu}$ and taking the $00$ and $ij$ components of the former we find,

 \underline{Conservation Laws of  Generic Cosmological Perfect Hyperfluid}:
		\beq
	\dot{\rho}_{c}+(n-1)H(\rho_{c}+p_{c})	=-\frac{1}{2}u^{\mu}u^{\nu}(\chi R_{\mu\nu}+\psi \breve{R}_{\mu\nu})+\frac{1}{2}(\rho_{c}-\rho)C+\frac{1}{2}(p_{c}-p)(n-1)A \label{cont}
	\eeq	
	\beq
	\dot{\phi}+(n-1)H \phi +H(\chi +\psi) +\psi X- \chi Y=2 (p_{c}-p) 	 \label{hyper1}
	\eeq
	\beq
	\dot{\omega}+(n-1)H(\chi+\psi+\omega)+(n-1)( \psi X-\chi Y)=2(\rho_{c}-\rho) \label{hyper2}
	\eeq
	Additionally, one could take the trace of ($\ref{conl2}$) to arrive at
	\beq
	(n-1)\dot{\phi}-\dot{\omega}+(n-1)H \Big[ (n-1)\phi-\omega \Big]=2\Big[ (\rho-\rho_{c})-(n-1)(p-p_{c})  \Big] \label{dil}
	\eeq
	However, as expected, this gives no further information since it is easily seen that the latter is equal to (n-1)(\ref{hyper1})-(\ref{hyper2}). Therefore, as we have already mentioned, the full dynamics of the Perfect Hyperfluid is contained in the three equations (\ref{cont}), (\ref{hyper1}) and (\ref{hyper2}). We should emphasize again that the latter equations are fairly general and hold true regardless of the equations of state to be imposed on the hyperfluid variables. For any Metric-Affine Cosmology, the evolution equations for the sources are the aforementioned three. 
	
	Note that when the hyperfluid is not hypermomentum preserving a direct physical meaning can be given to both $\phi$ and $\omega$. Indeed, as we see from the two hypermomentum equations (\ref{hyper1}) and (\ref{hyper2}) by solving with respect to the net density and pressure ($\rho_{c}, p_{c}$) the time derivatives $\dot{\phi}$ and $\dot{\omega}$ have respectively the dimensions of density and pressure. Therefore $\phi$ contributes to the fluid's density, while $\omega$ modifies its pressure.
	
Now, let us turn back to the generalized continuity equation. By adding a zero at the right-hand side of (\ref{cont}) we may bring it in the alternative form
	\beq
\dot{\rho}_{c}+(n-1)H(\rho_{c}+p_{c})	=\frac{(\psi-\chi)}{2}R_{\mu\nu}u^{\mu}u^{\nu}-\frac{\psi}{2}(R_{\mu\nu}+\breve{R}_{\mu\nu})u^{\mu}u^{\nu}+\frac{1}{2}(\rho_{c}-\rho)C+\frac{1}{2}(p_{c}-p)(n-1)A \label{cont3}
\eeq
The advantage of the latter is that it separates out the term $R_{\mu\nu}u^{\mu}u^{\nu}$ which can always be readily computed for any Theory by contracting the metric field equations with $u^{\mu}u^{\nu}$. In addition, there appears the combination $(R_{\mu\nu}+\breve{R}_{\mu\nu})$ which measures the non-integrability of the geometry. In particular when
\beq
 R_{\mu\nu}+\breve{R}_{\mu\nu}=0
 \eeq
  the geometry is said to be integrable (see for instance \cite{BeltranJimenez:2020sih}). This is always the case for Weyl-Cartan geometries. In fact, this term can be shown to be equal to (see ref. \cite{Iosifidis:2021crj})
	\begin{gather}
(R_{\mu\nu}+	\breve{R}_{\mu\nu})=\Big[ (\dot{X}+\dot{Y})+H (X+Y)+2H (Z+V) +(X-Y)(Z+V) \Big]\Big(h_{\mu\nu}+(n-1)u_{\mu}u_{\nu}\Big)
\end{gather}
that is (using also (\ref{dv}))
\beq
(R_{\mu\nu}+	\breve{R}_{\mu\nu})u^{\mu}u^{\nu}=\frac{(n-1)}{2} (\dot{B}+H B) +(n-1)(A+C)\Big( H-Y+\frac{B}{4}\Big) \label{RR}
\eeq
from which we see that indeed for Weyl-Cartan geometries (i.e. $B=0$, $A+C=0$) this quantity vanishes identically. Furthermore, let us note that, quite remarkably, the parenthesis in the second term of the right-hand side of the above has a clear physical interpretation, namely it is the generalized $n-1$ divergence of the velocity field. Indeed, since we now have spacetime non-metricity there are two ways to define the spatial (n-1) divergence of the spatial velocity $u^{i}$. These are
\beq
\nabla_{i}u^{i}=(n-1)(H-Y)
\eeq
and
\beq
g^{ij}\nabla_{i}u_{j}=(n-1)(H+X)
\eeq
Of course, for vanishing non-metricity these two coincide. It is therefore quite natural, when non-metricity is present, to define the spatial divergence as the mean value of the above two, namely\footnote{Remarkably, this is equal to the $n$-velocity mean divergence as well since a trivial calculation reveals $\frac{1}{2}(\nabla_{\mu}u^{\mu}+g^{\mu\nu}\nabla_{\mu}u_{\nu})\equiv  (\nabla \cdot u)$. In addition we see that the variable responsible for the non-equivalence of the two spatial divergencies is the $B$ part of non-metricity. In particular we have that $\nabla_{i}u^{i}-g^{ij}\nabla_{i}u_{j} =-\frac{1}{2}(n-1)B$ and thus the two coincide only for $B=0$. For instance this is always true for Weyl-Cartan geometries.}

\underline{Non-Riemannian Spatial Divergence:}

\beq
(\nabla \cdot u):=\frac{\nabla_{i}u^{i}+g^{ij}\nabla_{i}u_{j}}{2}  \label{gendiv}
\eeq 
This  serves as a well motivated generalization of the spatial divergence  in Non-Riemannian Geometries. For the above Cosmological setting, it takes the specific form
\beq
(\nabla \cdot u)=(n-1)\Big( H-Y+\frac{B}{4}\Big)  \label{nablau2}
\eeq
as stated. We shall therefore call a hyperfluid $\underline{incompresable}$ if $(\nabla \cdot u)=0$. With the above definition and using (\ref{RR}) we may express the modified continuity equation (\ref{cont3}) in the alternative form
	\beq
\dot{\rho}_{c}+(n-1)H(\rho_{c}+p_{c})	=\frac{(\psi-\chi)}{2}R_{\mu\nu}u^{\mu}u^{\nu}-\frac{\psi}{2}\Big( (\dot{B}+H B)+(A+C)   (\nabla \cdot u) \Big)+ \frac{1}{2}(\rho_{c}-\rho)C+\frac{1}{2}(p_{c}-p)(n-1)A \label{cont4}
\eeq
It is also worth stressing out that for the case of totally symmetric traceless non-metricity, the above velocity divergence is independent of non-metricity since in this case, using equations (\ref{totsym}), it is easy to show that $(\nabla \cdot u)=(n-1)( H-2 \Phi)$.
	
	\subsection{Hyperfluids Classification}
	
	As we have already discussed, the hypermomentum current splits into its three irreducible pieces of shear, spin and dilation. It is therefore interesting to study special cases of hyperfluids containing only one of the above parts each time. In particular, below we classify and give the form that the conservation laws take for the cases of pure shear, pure spin and pure dilation hyperfluids.
	
	\subsubsection{Pure Shear Hyperfluid}
	In this case, the vanishing of spin and dilation currents imply
	\beq
	\psi=\chi \;\;, \;\; \zeta=0\;\;, \;\; \omega=(n-1)\phi
	\eeq
	and the hypermomentum is given by
	\beq
	\Delta_{\alpha\mu\nu}=\breve{\Delta}_{\alpha\mu\nu}=\Big[ h_{\alpha\mu}+(n-1)u_{\alpha}u_{\mu}\Big] \phi + 2 \psi u_{(\mu}h_{\nu)\alpha}
	\eeq
	The conservation laws take the form
		\beq
	\dot{\rho}_{c}+(n-1)H(\rho_{c}+p_{c})	=-\frac{\psi}{2}( R_{\mu\nu}+ \breve{R}_{\mu\nu})u^{\mu}u^{\nu}+\frac{1}{2}(\rho_{c}-\rho)(A+C) \label{r}
	\eeq
	\beq
	\dot{\phi}+H \Big( (n-1) \phi +2\psi \Big) +\psi (X-Y)=2 (p_{c}-p) 	 \label{phi}
	\eeq
	\beq
	\dot{\omega}+(n-1)H(\omega+ 2\psi)+(n-1)\psi ( X-Y)=2(\rho_{c}-\rho)  \label{omega}
	\eeq
	\beq
	(\rho_{c}-\rho) =(n-1)(p_{c}-p) 
	\eeq
	In the above, the first one is the modified continuity equation, the second and third equations are the conservation laws for the hypermomentum degrees of freedom and the last relation is a consequence of the vanishing dilation. Note that for Weyl non-metricity (i.e. $A=-C$ $and$ $B=0$) it holds that\footnote{This is so because in the Cosmological expression for this combination, there appear terms proportional (see ref \cite{Iosifidis:2021iuw}) to $B$ and $(A+C)$ which as mentioned both vanish for Weyl non-metricity.} $ R_{\mu\nu}+ \breve{R}_{\mu\nu}=0$ and in this instance both terms on the right-hand side of (\ref{r}) drop out and one obtains the usual form of the continuity equation. In particular, a \underline{Pure Shear Weyl Hyperfluid} obeys the usual continuity equation
	\beq
		\dot{\rho}_{c}+(n-1)H(\rho_{c}+p_{c})=0
	\eeq
	It is also worth mentioning that for such a fluid only the continuity equation for the density decouples from the hypermomentum variables, while the dynamics of the latter involves the perfect fluid contributions of pressure and density as it seen from (\ref{phi}) and (\ref{omega}). However, this is not the case if the hyperfluid is hypermomentum preserving, in which case the right-hand sides of both of the aforementioned equations vanish. More concretely, the \underline{Pure Shear, Hypermomentum preserving hyperfluid} is governed by the conservation laws 
	
		\beq
	\dot{\rho}+(n-1)H(\rho+p)	=-\frac{\psi}{2}( R_{\mu\nu}+ \breve{R}_{\mu\nu})u^{\mu}u^{\nu},\; \; p_{c}=p \;, \;\; \rho_{c}=\rho 
	\eeq
	\beq
	\dot{\phi}+H \Big( (n-1) \phi +2\psi \Big) +\psi (X-Y)=0
	\eeq
	\beq
	\dot{\omega}+(n-1)H(\omega+ 2\psi)+(n-1)\psi ( X-Y)=0
	\eeq

	Notice that now the decoupling is reversed, namely the continuity equation for the density receives contributions from hypermomentum while the conservation laws for the hypermomentum variables completely decouple from $\rho_{c}$ and $p_{c}$. It is then natural to ask, when do all both os this sets of conservation laws completely decouple from one another. As we shall see, in what follows, this happens for the case of a pure dilation and hypermomentum preserving hyperfluid.

 It is also worth mentioning that when the Gravitational Lagrangian consists only of the Einstein-Hilbert term (i.e. the Ricci scalar) then by employing the connection field equations
 \begin{gather}
     -\nabla_{\lambda}(\sqrt{-g}g^{\mu\nu})+\nabla_{\sigma}(\sqrt{-g}g^{\mu\sigma})\delta^{\nu}_{\lambda} 
+\sqrt{-g}(S_{\lambda}g^{\mu\nu}-S^{\mu}\delta_{\lambda}^{\nu}+g^{\mu\sigma}S_{\sigma\lambda}^{\;\;\;\;\nu})=\kappa \Delta_{\lambda}^{\;\;\mu\nu}
 \end{gather}
 and the identity \cite{Iosifidis:2020gth}
 \begin{gather}
\hat{\nabla}_{\nu}\Big( -\nabla_{\lambda}(\sqrt{-g}g^{\mu\nu})+\nabla_{\sigma}(\sqrt{-g}g^{\mu\sigma})\delta^{\nu}_{\lambda} 
+\sqrt{-g}(S_{\lambda}g^{\mu\nu}-S^{\mu}\delta_{\lambda}^{\nu}+g^{\mu\sigma}S_{\sigma\lambda}^{\;\;\;\;\nu}) \Big)=\sqrt{-g}g^{\mu\nu}(\breve{R}_{\nu\lambda}+R_{\lambda\nu})
\end{gather}
it follows that
\beq
\kappa \hat{\nabla}_{\nu}\Big(    \sqrt{-g}\Delta_{\lambda}^{\;\;\;\mu\nu}\Big)=\sqrt{-g}g^{\mu\nu}(\breve{R}_{\nu\lambda}+R_{\lambda\nu})\;\; , \;\; \hat{\nabla}_{\nu}=2 S_{\nu}-\nabla_{\nu} \label{DD2}
\eeq
	Therefore, for the hypermomentum preserving case it holds that
\beq
\Big(R_{\mu\nu}+ \breve{R}_{\mu\nu}\Big)u^{\mu}u^{\nu}=0
\eeq
 and as a result the continuity equation, for the hypermomentum preserving pure shear fluid, is again given by its usual form being free from any hypermomentum contributions. In fact the above result is not restricted only for $\mathcal{L}_{G}=R$ but it holds true for general Metric-Affine $f(R)$ Theories.

	\subsubsection{Pure Spin}
 Now, imposing vanishing shear and dilation at the same time, it follows that
	 \beq
	 \phi=0 \;\;, \;\; \omega=0\;\;, \;\; \psi=-\chi
	 \eeq
	with hypermomentum
	\beq
	\Delta_{\alpha\mu\nu}=	\Delta_{[\alpha\mu]\nu}=2 \psi u_{[\alpha}h_{\mu]\nu}+\delta_{n,4}\varepsilon_{\alpha\mu\nu\kappa}u^{\kappa}\zeta
	\eeq
	The associated conservation laws are in this case
	\beq
	\dot{\rho}_{c}+(n-1)H(\rho_{c}+p_{c})	=+\frac{\psi}{2}( R_{\mu\nu}- \breve{R}_{\mu\nu})u^{\mu}u^{\nu}+\frac{1}{2}(\rho_{c}-\rho)(A+C) \label{r2}
	\eeq
		\beq
	 +\psi (X-Y)=2 (p_{c}-p) 	\label{r}
	\eeq
	\beq
(n-1)\psi ( X-Y)=2(\rho_{c}-\rho) \label{p}
	\eeq
	\beq
	(\rho_{c}-\rho) =(n-1)(p_{c}-p) 
	\eeq
	It is worth stressing out that in this instance, the conservation laws for the hypermomentum variables become simple algebraic relations between the latter  and the effective pressure and density of the hyperfluid. Interestingly, in a hypermomentum dominated era it holds that $p \approx 0$, $\rho \approx 0$ and consequently
	\beq
	p_{c}\approx \frac{\psi}{2}(X+Y) =\frac{1}{4}\psi B
	\eeq
	\beq
	\rho_{c}\approx (n-1)\frac{\psi}{2}(X+Y)=\frac{(n-1)}{4} \psi B
	\eeq
	and then the above continuity equations gives dynamics to $\psi$ as well. Note that the $B$ part of non-metricity is all essential here in order to get a non-trivial dynamics. It is also noteworthy that for a hypermomentum preserving hyperfluid the conservation laws for the hypermomentum variables trivializes and therefore there is no such thing as a hypermomentum preserving pure spin fluid. Finally, let us observe that the conservation laws have no explicit dependence on the $\zeta$-part of spin. They do however depend implicitly on the latter since the torsion and non-metricity variables along with the Ricci and co-Ricci tensor can depend on it. 

 More information about the form of (\ref{r}) and (\ref{p}) can be extracted if further dynamical characteristics are known.  
 For instance, for the wide class of quadratic (in torsion and non-metricity) MAG Theories all distortion variables are linearly related to the hypermomentum sources (see \cite{Iosifidis:2021bad}), here $\psi$, and consequently we find the net density and pressure
 \beq
\rho_{c}=\rho+\frac{(n-1)}{2}\gamma_{0}\psi^{2}
 \eeq
\beq
p_{c}=p+\frac{1}{2}\gamma_{0}\psi^{2}
\eeq
where $\gamma_{0}$ depends on the parameters of the quadratic Theory. We see therefore that the spin part of hypermomentum modifies the density and pressure of the fluid in a non-trivial way.

		\subsubsection{Pure Dilation}
		Setting to zero both the shear and spin currents  we obtain
		\beq
		\psi=0\;\;, \;\; \chi=0\;\;, \;\; \zeta=0 \;\;, \;\; \omega=-\phi
		\eeq
	and as a result
	\beq
	\Delta_{\alpha\mu\nu}=\phi g_{\alpha\mu}u_{\nu}
	\eeq
	with the conservation laws becoming,

 \underline{Pure Dilation Hyperfluid}:
		\beq
	\dot{\rho}_{c}+(n-1)H(\rho_{c}+p_{c})	=\frac{1}{2}(\rho_{c}-\rho)\Big( C-(n-1)A \Big) \label{co}
	\eeq	
	\beq
	\dot{\phi}+(n-1)H \phi =2 (p_{c}-p) 	 
	\eeq
	\beq
	\dot{\omega}+(n-1)H\omega=2(\rho_{c}-\rho) 
	\eeq
	\beq
	(p_{c}-p)=-(\rho_{c}-\rho) 
	\eeq
	Observe now that  the continuity equation (\ref{co}) does not receive hypermomentum contributions if, obviously, the hyperfluid is hypermomentum preserving or  if, quite intriguingly, the non-metricity is of the totally symmetric traceless type as given by eq. (\ref{qtraceless}). The former restricts the form of the fluid while the latter restricts the geometry. In either case the continuity equation acquires its usual form, evolving as if the hypermomentum was not there. In particular, if the hyperfluid is hypermomentum preserving both the Perfect Fluid characteristics and the Hypermomentum variables totally decouple from one another, and evolve as follows,

 \underline{Pure Dilation Hypermomentum Preserving Hyperfluid}:
 
	\beq
	\dot{\rho}+(n-1)H(\rho+p)	=0 \;\;, \;\; \rho_{c}=\rho\;, \; p_{c}=p
	\eeq	
	\beq
	\dot{\phi}+(n-1)H \phi = 0 \;\;, \;\; \omega=-\phi \label{dustlike}
	\eeq
Thus, the usual continuity equation is recovered for the Perfect Fluid component and, in addition, the dust-like evolution equation for $\phi$ immediately integrates to
\beq
\phi \propto \frac{1}{a^{n-1}} 
\eeq
And the usual $\rho \propto 1/a^{(n-1)(1+w)}$ can also be obtained for the former, if the barotropic equation of state $p=w \rho$ is assumed. Let us note that even though for $n=4$ the above form  for $\phi$ is similar to that of dust (i.e. $\propto 1/a^{3}$) in the modified Friedmann equations, there appear $\phi^{2}$ terms meaning that the final effect is similar to that of a stiff matter component in this case \cite{Iosifidis:2020upr}.

	\subsection{Alternative forms of the conservation laws in terms of the Physical Variables}
	
	As we can see from the hypermomentum decomposition (\ref{hypsplit}) the fields $\psi, \chi$ etc.  have no direct physical interpretation just by themselves, but rather specific combinations of them correspond to the spin, dilation and shear of matter. In particular, defining 
	\beq
	\sigma:=\frac{(\psi-\chi)}{2}\;\;, \;\;\zeta \;\;\;\;\;\;\; (spin)
	\eeq 
	\begin{equation}
\Delta :=(n-1) \phi-\omega \;\;\;\;\;\;\; (dilation)
	\end{equation}
	\beq
\Sigma_{1}:=	\frac{(\psi+\chi)}{2} \;\;, \;\; \Sigma_{2}:=\frac{(\phi+\omega)}{n} \;\;\;\;\;\;\; (shear)
	\eeq
	In the above, the first one is the first combination of spin, in the second line the dilation and $\Sigma_{1}, \Sigma_{2}$ correspond to the two parts of shear. With these redefinitions the three sources of the hypermomentum take the form
		\beq
	\Delta_{[\alpha\mu]\nu}=2 \sigma u_{[\alpha}h_{\mu]\nu}+\delta_{n,4}\epsilon_{\alpha\mu\nu\kappa}u^{\kappa}\zeta
	\eeq
	\beq
	D_{\nu}:=\Delta_{\alpha\mu\nu}g^{\alpha\mu}=\Delta u_{\nu}   \label{dil}
	\eeq
	\beq
	\breve{\Delta}_{\alpha\mu\nu}=\Delta_{(\alpha\mu)\nu}-\frac{1}{n}g_{\alpha\mu}D_{\nu} =\Sigma_{2}\Big[ h_{\alpha\mu}+(n-1)u_{\alpha}u_{\mu} \Big] u_{\nu} +2 \Sigma_{1} u_{(\mu}h_{\alpha)\nu}
	\eeq
	The inverse relations read
	\beq
\Sigma_{1}+\sigma=\psi \;\;, \;\; 	\Sigma_{1}-\sigma=\chi \;\;, \;\; \phi=\Sigma_{2}+\frac{1}{n}\Delta \;\;, \;\; \omega=(n-1)\Sigma_{2}-\frac{1}{n}\Delta 
	\eeq
	Using these we can bring the conservation laws of the hyperfluid in the alternative form
		\beq
	\dot{\rho}_{c}+(n-1)H(\rho_{c}+p_{c})	=\sigma R_{\mu\nu}u^{\mu}u^{\nu} +(\Sigma_{1}+\sigma)( R_{\mu\nu}+ \breve{R}_{\mu\nu})u^{\mu}u^{\nu}+\frac{1}{2}(\rho_{c}-\rho)C+\frac{1}{2}(p_{c}-p)(n-1)A 
	\eeq
\beq
\dot{\Delta}+(n-1) H \Delta =2 \Big[ (n-1)(p_{c}-p)-(\rho_{c}-\rho) \Big]
\eeq
\beq
\dot{\Sigma}_{2}+(n-1)H \Sigma_{2}+ 2\Sigma_{1}(H-Y)+\frac{B}{2}(\Sigma_{1}+\sigma)=\frac{2}{n}\Big[ (p_{c}-p)+(\rho_{c}-\rho) \Big]
\eeq	
	The first one being the modified continuity equation, the second  governs the evolution of the dilation current and the last one is the evolution equation for the shear part. It is worth stressing out that in the continuity equation, the spin part directly couples to the usual  perfect fluid  	contributions (i.e. $p$ and $\rho$) as seen by the presence of the term $\sigma R_{\mu\nu}u^{\mu}u^{\nu}$. In addition, we see that one part of each  spin and shear ($\sigma, \Sigma_{1}$) couple to the non-integrability term. The dilation part involves independently from the rest of the hypermomentum variables and the evolution equation of the second part of shear contains contributions from the first part as well as from spin.
	
	Now, it is evident from the above equations that, in this picture, the dynamics of the hypermomentum degrees of freedom is contained in $\Sigma_{2}$ and $\Delta$, namely in the second shear part and in dilation. Therefore, following the perfect fluid paradigm, it is quite reasonable to supplement the system with equations of state relating the rest of the hypermomentum variables to these two fields in order to close the system of equations\footnote{Again, this is really no different from the case of the classical perfect fluid case where we have the continuity equation ... giving us the evolution of the density .. and one has to impose an equation of state p($\rho$), usually in the barotropic form $p=w \rho$ in order to close the system of equations.}. As a result, we have the following set of equations of state\footnote{These will be true given that $\Sigma_{2}$ and $\Delta$ are not both vanishing at the same time.}
	\beq
	\Sigma_{1}=w_{1}\Sigma_{2}+w_{2}\Delta \;\;, \;\; \sigma=w_{3}\Sigma_{2}+w_{2}\Delta \;\;, \;\;\zeta =w_{5}\Sigma_{2}+w_{6}\Delta  \label{eos}
	\eeq
	With these at hand, and recalling $(\ref{RR})$ and $(\ref{nablau2})$ we readily arrive at
		\beq
	\dot{\rho}_{c}+(n-1)H(\rho_{c}+p_{c})	=\sigma R_{\mu\nu}u^{\mu}u^{\nu}-\frac{1}{2}(w_{13}\Sigma_{2}+w_{24}\Delta )\Big[ \frac{(n-1)}{2}(\dot{B}+H B)+(A+C)(\nabla \cdot u) \Big]  +\frac{1}{2}\bar{\rho}C+\frac{1}{2}\bar{p}(n-1)A 
	\eeq
\beq
\dot{\Delta}+(n-1) H \Delta =2 \Big[ (n-1)\bar{p}-\bar{\rho} \Big]
\eeq
\beq
\dot{\Sigma}_{2}+(n-1)H \Sigma_{2}+ \frac{2}{(n-1)}(w_{1}\Sigma_{2}+w_{2}\Delta)(\nabla \cdot u)+\frac{B}{2}(w_{13}\Sigma_{2}+w_{24}\Delta )=\frac{2}{n}( \bar{p}+\bar{\rho} )
\eeq	
	where we have set $w_{13}=w_{1}+w_{3}$, $w_{24}=w_{2}+w_{4}$ and also $\bar{\rho}=(\rho_{c}-\rho)$ and $\bar{p}=p_{c}-p$. Then, considering also an equation of state $\bar{p}=\bar{w}\bar{\rho}$ and also the usual one $p=w \rho$ the system of equations is closed since we have $8$ equations among the matter fields (3 for perfect fluid components + 5 for hypermomentum) and also for a given theory the first Friedmann equation is known yielding in total a system of 9  unknowns with 9 equations.  Note that the pure spin hyperfluid must be studied separately since the above analysis is valid only when $\Sigma_{2}$ and $\sigma$ are non-vanishing at the same time. This is of course not true for the pure spin case. In this instance, for the evolution equations for shear and dilation it follows that
	\beq
	\bar{p}+\bar{\rho}=\frac{n \sigma B}{4} \;\;, \;\;(n-1)\bar{p}=\bar{\rho}
	\eeq
	which is trivially solved giving
	\beq
	\bar{p}=\frac{\sigma B}{4}\;\;, \;\; \bar{\rho}=(n-1)\frac{\sigma B}{4}
	\eeq
	Substituting these into the modified continuity equation, the latter takes the form
		\beq
	\dot{\rho}_{c}+(n-1)H(\rho_{c}+p_{c})	=\sigma R_{\mu\nu}u^{\mu}u^{\nu} +\frac{(n-1)}{8}\sigma B (A+C)
	\eeq
	Interestingly, the last term expresses a coupling between spin and non-metricity. This is vanishing for an extended Weyl-non-metricity or for $B=0$. Note also that the condition $B=0$ implies that the fluid is necessarily hypermomentum preserving (i.e. $\rho_{c}=\rho$ and $p_{c}=p$). As we saw previously, the non-vanishing of $B$ also implies that the two possible definitions of the velocity divergence do not coincide. We see therefore that the $B$ degree of freedom of non-metricity is responsible for complications in many instances.

	\section{Incompressible Hyperfluids }
	
	It is well known from continuum mechanics (see for instance) that in describing the continuum motion one needs to impose further equations among the variables in order to obtain a complete closed system. These additional required data often comes in the form of constitutive equations. The latter depend on the constitution of the material and as such are not universal but rather change from one medium to another. However, a wide variety of fluids are to a good approximation what we call incompressible, in the sense that their volume remains unchanged during the motion. In Euclidean space, this is mathematically expressed by the vanishing of the divergence of the spatial velocity field, namely
	\beq
	\vec{\nabla} \cdot \vec{u}=0
	\eeq
	Following this definition, in a non-Riemannian space, we shall also call a hyperfluid incompressible if the generalized spatial divergence, as we have defined it in (\ref{gendiv}) is vanishing,

	\underline{Incompressible Hyperfluid}:
	\beq
(\nabla \cdot u):=\frac{\nabla_{i}u^{i}+g^{ij}\nabla_{i}u_{j}}{2}=0 \label{nablau0}
\eeq 
The latter is true for  hyperfluids that are not necessary homogeneous. If we demand also homogeneity  we may use (\ref{nablau2}) and then the above relation implies,

		\underline{Incompressible Cosmological Hyperfluid}:
\beq
	 H-Y+\frac{B}{4}=0 \label{HB}
\eeq
	Under this condition, the conservation laws become
		\beq
	\dot{\rho}_{c}+(n-1)H(\rho_{c}+p_{c})	=\sigma R_{\mu\nu}u^{\mu}u^{\nu}-\frac{1}{2}(\Sigma_{1}+\sigma) \frac{(n-1)}{2}(\dot{B}+H B)+  +\frac{1}{2}\bar{\rho}C+\frac{1}{2}\bar{p}(n-1)A 
	\eeq
\beq
\dot{\Delta}+(n-1) H \Delta =2 \Big[ (n-1)\bar{p}-\bar{\rho} \Big]
\eeq
\beq
\dot{\Sigma}_{2}+(n-1)H \Sigma_{2}+\sigma \frac{B}{2}=\frac{2}{n}( \bar{p}+\bar{\rho} )
\eeq	
These are the conservation laws for incompressible Cosmological Hyperfluids. In addition, in this case eq. (\ref{HB}) is immediately integrated to provide the integral of motion
\beq
a(t)=c e^{\int \Big( Y-\frac{B}{4}\Big) dt }\;\;, \;\; c=const.
\eeq
Of course this should not be seen as a solution for the scale factor since the various quantities appearing on the right-hand side depend also on $a$. All we can infer from the above is an integral of motion constraining the hyperfluid evolution a certain way when the latter is incompressible.	 

By enlarging the number of constitutive equations it is possible to obtain certain kinematical characteristics of the specific fluids that  fall into such categories. For instance for a hyperfluid that is i) Hypermomentum Preserving ($\rho_{c}=\rho$, $p_{c}=p$), ii) Spinless ($\sigma=0$, $\zeta=0$) and  iii) Incompressible ($\nabla \cdot u =0$) the conservation laws boil down to
\beq
\dot{\rho}+(n-1)(1+w)H \rho=-\frac{(n-1)}{4}\Sigma_{1}(\dot{B}+H B) \label{r}
\eeq
\beq
\dot{\Delta}+(n-1) H \Delta =0
\eeq
\beq
\dot{\Sigma}_{2}+(n-1)H \Sigma_{2}=0
\eeq	
where we have also assumed the usual barotropic equation of state $p=w \rho$ among the perfect fluid variables of the hyperfluid. Now, the last two relations are trivially integrated to give
\beq
\Sigma_{2}\propto \Delta \propto \frac{1}{a^{(n-1)}}
\eeq
Note that the last result is kinematic, that is, it is independent of the gravitational action that one considers. Is is also interesting to obtain some dynamical characteristics for a class of Theories. For instance, for gravitational actions that contain the usual Einstein-Hilbert term and quadratic invariants of torsion and non-metricity \cite{Iosifidis:2021tvx}, the torsion and no-metricity variables are linearly related to the hypermomentum sources (see \cite{Iosifidis:2021tvx}). In this case $B$ will be linear in $\Delta, \Sigma_{2}$, which when combined with the above results implies that\footnote{Using that $\Sigma_{2}=C_{0}/a^{(n-1)}$ and $\Delta=\tilde{C}_{0}/a^{(n-1)}$.}
\beq
B=\frac{B_{0}}{a^{(n-1)}}
\eeq
for some constant $B_{0}$.	Then, substituting this result back in $(\ref{r})$ and also considering the barotropic equations of state $p=w \rho$ and $\Sigma_{1}=w_{\Sigma}\Sigma_{2}$, we obtain
\beq
\dot{\rho}+(n-1)(1+w)H \rho=\lambda_{0}B \dot{B} \;\;, \;\; \lambda_{0}=\frac{C_{0}w_{\Sigma}(n-2)}{2 B_{0}}
\eeq
The last is then trivially integrated to give
\beq
\rho=\frac{\lambda_{1}}{a^{2(n-1)}}+\frac{C}{a^{(n-1)(1+w)}}
\eeq
that is, for our physical $n=4$ Cosmology	
\beq
\rho=\frac{\lambda_{1}}{a^{6}}+\frac{C}{a^{3(1+w)}}
\eeq
where $\lambda_{1}=-(n-1)\lambda_{0}B_{0}^{2}$ and $C$ is an integration constant. From the latter we can infer a quite interesting conclusion.  We see that regardless the value of the barotropic index\footnote{Of course on the premise that $w_{\Sigma}\neq 0$ since otherwise the continuity equation is completely decoupled from any hypermomentum contribution in this case.} $w_{\Sigma}$ there is an extra non-Riemannian modification on the fluid's density $\rho$ which mimics always that of  a stiff fluid matter component ($\propto 1/a^{6}$).  Conversely, we see that in the generalized non-Riemannian setup, a stiff matter fluid is not so exotic but rather it is just a specific manifestation of the hypermomentum sources of matter. In other words, there is no need to arbitrary impose a stiff matter barotropic equation of state, the latter arises naturally in a Metric-Affine framework . Additionally, from the above equation we see that we have also the usual perfect fluid term $\frac{C}{a^{3(1+w)}}$ as expected. When the non-Riemannian degrees of freedom are switched-off the density evolves in the usual  manner ($\propto a^{3(1+w)}$), as it should.

Let us note that the same behaviour appears even if we relax the condition iii) Incompressability and instead consider a Weyl geometry. Indeed, since the second term of (\ref{RR}) vanishes in this case as well\footnote{Recall that the combination $A+C$ vanishes for Weyl geometries.} similar results to the above follow immediately.

	\section{An Example: Scalar Field Coupled to the Connection}

		In order to give a concrete example of how non-vanishing hypermomentum contributions come into the gave we shall consider a simple example of a scalar field coupled to the connection. We start off with the usual matter action of a single scalar field $\vartheta$ as given by
		\beq
	S_{M}^{(0)}=\int \sqrt{-g}d^{n}x	\mathcal{L}^{(0)}=\int \sqrt{-g}d^{n}x\big[-\frac{1}{2}\partial_{\mu}\vartheta\partial^{\mu}\vartheta -V(\vartheta) \Big]
		\eeq
Now, the simplest couplings we can construct among the scalar field and the connection are derivative couplings of the fields with the torsion and non-metricity vectors (\cite{shimada2019metric,Iosifidis:2020gth}). Including these couplings our final matter action reads
\beq
	S_{M}=\int \sqrt{-g}d^{n}x	\mathcal{L}=\int \sqrt{-g}d^{n}x\Big[-\frac{1}{2}\partial_{\mu}\vartheta\partial^{\mu}\vartheta -V(\vartheta) + (a_{1}S_{\mu}+a_{2}Q_{\mu}+a_{3}q_{\mu}+b_{1}t_{\mu})\vartheta \partial^{\mu}\vartheta\Big]
		\eeq
where we have introduced $\vartheta \partial_{\mu} \vartheta$ couplings instead of just $\partial_{\mu} \vartheta$ in order for the parameters $a_{i}$, $i=1,...,4$ and $b_{1}$ to be dimensionless. The hypermomentum associated to this action, reads

\beq
\Delta_{\alpha\mu\nu}=-(a_{1}+2 a_{3})g_{\alpha\nu}\vartheta \partial_{\mu}\vartheta+(a_{1}-4 a_{2})g_{\alpha\mu}\vartheta \partial_{\nu}\vartheta-2 a_{3}g_{\mu\nu}\vartheta \partial_{\nu}\vartheta-b_{1}\varepsilon_{\alpha\mu\nu}^{\;\;\;\;\;\;\lambda}\vartheta \partial_{\lambda}\vartheta
\eeq
	Then, assuming  a  homogeneous and isotropic Cosmological setting (i.e FLRW), using $h_{\mu\nu}=g_{\mu\nu}+u_{\mu}u_{\nu}$ and comparing the above expression with (\ref{Dform}) we trivially find
	\beq
	\phi=(4 a_{3}-a_{1})\vartheta\dot{\vartheta}\;\;, \;\; \chi=(a_{1}+2 a_{3})\vartheta\dot{\vartheta}\;\;, \;\; \psi=2 a_{3}\vartheta\dot{\vartheta}
	\eeq
	\beq
	\omega=-4(a_{2}+a_{3})\vartheta\dot{\vartheta}\;\;, \;\; \zeta=b_{1}\vartheta\dot{\vartheta}
	\eeq
	Therefore for the three constituents of the hypermomentum source, we have:
	
	\begin{enumerate}
	    \item \underline{Spin}
	    
	 We see that the first spin source is given by
	 \beq
	 \sigma:=\frac{1}{2}(\psi-\chi)=-\frac{1}{2}a_{1}\vartheta  \dot{\vartheta}
  \eeq
and the second
\beq
  \zeta=b_{1}\vartheta\dot{\vartheta}
	 \eeq
	 There is no surprise here since the totally antisymmetric part of hypermomentum is related to the component of $t^{\mu}$ which is extracted by the totally antisymmetric part of torsion. Generally, we observe that in order that for a spin part, derivative couplings between the scalar field and the torsion vectors are all essential. As a result, torsion-scalar interactions are necessary in the Lagrangian $\mathcal{L}_{M}$ to guarantee a non-vanishing spin part for hypermomentum. In addition, we see that at this order (i.e. with these couplings alone) the spin part has no relation to non-metricity. 
	 \item \underline{Dilation}
	 
	 The dilation part reads
	 \beq
	\Delta:= (n-1)\phi-\omega =\Big[ (a_{2}+n a_{3})-(n-1)a_{1}\Big] \vartheta \dot{\vartheta}
	 \eeq
	 Thus, we see that dilation can be generated by either torsion-scalar or non-metricity-scalar derivative interactions, or both at the same time. Note that the dilation part can just as well be associated to torsion and not necessarily to non-metricity exclusively.

	 \item \underline{Shear}

  The two parts of shear are given by
	 \beq
	 \Sigma_{1}:=\frac{1}{2}(\psi+\chi)=\frac{1}{2}(a_{1}+4 a_{3}) \vartheta \dot{\vartheta}\;\;, \;\; \Sigma_{2}:=\frac{1}{n}(\phi+\omega)=-\frac{1}{n}(a_{1}+4 a_{2}) \vartheta \dot{\vartheta}
	 \eeq
	 Here we see that the first part of shear $(\Sigma_{1})$ is related to the couplings $Q_{\mu}\partial^{\mu}\phi$ and $S_{\mu}\partial^{\mu}\phi$ while the second one is due to $q_{\mu}\partial^{\mu}\phi$ and $S_{\mu}\partial^{\mu}\phi$. Note that the torsion coupling generates both of the shear parts while the non-metricity vectors share the two components of shear.
	 
	\end{enumerate}
	
	From the above example, it is worth mentioning that all parts of hypermomentum have an association to torsion (namely the coupling to $S_{\mu}$ generates all parts of hypermomentum) while non-metricity is associated only to the dilation and shear parts. Let us also stress out that in  the above example we have the proportionality relations $\sigma\propto \zeta \propto \Delta \propto \Sigma_{1}\propto \Sigma_{2} \propto \vartheta \dot{\vartheta}$ which further motivate the introduction of barotropic equations of state, as in (\ref{eos}), among the Hyperfluid variables.

	\section{A first Integral}
	We shall now obtain a first integral for non-Riemannian Cosmologies relating the scale factor (metric) with the distortion variables (affine connection). The expression we derive below is kinematic in the sense that its form is independent of the Gravitational action one considers. We have the following.
	
	\begin{theorem}
	Consider an n-dimensional non-Riemannian FLRW spacetime.Then for any flow, there always exists the integral  of motion
	\beq
a(t)=e^{\int Y dt}\left[ c_{2}+\int e^{\int (Z+V-Y)dt}\Big( c_{1}-\frac{1}{(n-1)}\int R_{\mu\nu}u^{\mu}u^{\nu} a e^{-\int (Z+V)dt} dt \Big) dt \right]	\label{ae}
	\eeq
	where $c_{1}$ and $c_{2}$ are integration constants.
		
  \begin{proof}
		We start from  the most general form of the acceleration equation  (aka Raychaudhuri eqn) with non-Riemannian effects included, which in this spacetime reads \cite{iosifidis2020cosmic}
			\begin{gather}
			\frac{\ddot{a}}{a}=-\frac{1}{(n-1)}R_{\mu\nu}u^{\mu}u^{\nu}+2\left( \frac{\dot{a}}{a} \right)\Phi +2\dot{\Phi} \nonumber \\
			+\left( \frac{\dot{a}}{a} \right)\left(A+\frac{C}{2}\right)+\frac{\dot{A}}{2}-\frac{A^{2}}{4}-\frac{1}{4}AC  
			-A\Phi-C \Phi\label{T}
					\end{gather}
					Firstly we use the set of equations (\ref{dv}) in order to express the torsion and non-metricity functions in terms of the distortion degrees of freedom, and the above recasts to
					\beq
					\frac{\ddot{a}}{a}=(\dot{Y}+H Y) +(Z+V)(H- Y)-\frac{1}{(n-1)}R_{\mu\nu}u^{\mu}u^{\nu}
					\eeq
				Now using the fact that $\ddot{a}/a=\dot{H}+H^{2}$ we observe that the later can be written as follows
				\beq
				\frac{d}{dt}(H-Y)+H(H-Y)=-\frac{1}{(n-1)}R_{\mu\nu}u^{\mu}u^{\nu}+(Z+V)(H-Y)
				\eeq	
				that is setting $\xi:=\nabla_{i}u^{i}=H-Y$, we have
				\beq
				\dot{\xi}+H \xi=-\frac{1}{(n-1)}R_{\mu\nu}u^{\mu}u^{\nu}+(Z+V)\xi
				\eeq
				then bringing $(Z+V)\xi$ to the left-hand side, multiplying by $e^{\int (H-Z-V) dt}$ and using the product rule of differentiation it follows that
    \beq
\frac{d}{dt} \Big[ e^{\int (H-Z-V) dt} \xi\Big]=-\frac{1}{(n-1)} R_{\mu\nu}u^{\mu}u^{\nu} e^{-\int (Z+V)dt} a 
    \eeq
	which integrates to\footnote{Also using the fact that $e^{\int H dt}=a$.}
\beq
\xi=\frac{1}{a}e^{\int (Z+V)dt}\left[ c_{1}-\int \frac{1}{(n-1)} R_{\mu\nu}u^{\mu}u^{\nu} e^{-\int (Z+V)dt} a dt \right]
\eeq
where $c_{1}$ is an integration constant. Then substituting back $\xi=H-Y$, multiplying through by the $e^{\int (H-Y) dt}$ and again noting the product rule it easily follows that
\beq
\frac{d}{dt} \Big[ e^{-\int Y dt} a\Big]=e^{\int (Z+V-Y)dt}\left[ c_{1}-\int \frac{1}{(n-1)} R_{\mu\nu}u^{\mu}u^{\nu} e^{-\int (Z+V)dt} a dt \right]
\eeq
which after another integration trivially yields the stated result
\beq
a(t)=e^{\int Y dt}\left[ c_{2}+\int e^{\int (Z+V-Y)dt}\Big( c_{1}-\frac{1}{(n-1)}\int R_{\mu\nu}u^{\mu}u^{\nu}  e^{-\int (Z+V) dt} a dt \Big) dt \right]	
	\eeq
			\end{proof}
	
	\end{theorem}
	\textbf{Comment $1$.} As already stressed out, the above expression is not the solution for the scale factor but merely a first integral of motion constraining the evolution of the latter together with rest of the hyperfluid variables. This is so because the functions $X,Y Z,V,W$ may as well contain an explicit dependence on the scale factor.

	\subsection{Application: Constant Non-Riemannian Background}
 As a direct application of the above result let us consider a generalized Ricci flat space, i.e. $R_{(\mu\nu)}=0$ and a constant torsion and non-metricity configuration: $A=A_{0}=constant$, $\Phi=\Phi_{0}$ etc. In this case an analytic expression for the scale factor is possible and we trivially find
 \beq
a(t)=e^{Y_{0}t}\Big[ c_{2}+\frac{c_{1}}{(Z_{0}+V_{0}-Y_{0})}e^{(Z_{0}+V_{0}-Y_{0})t}  \Big]\label{td}
 \eeq
	Interestingly,  for extended Weyl-Cartan geometries (i.e for non-metricity with $A+C=0$, $B\neq 0$) the above expression becomes $a(t)\propto c_{2}e^{Y_{0}t}+C_{1}$, with $C_{1}=const.$ which for $Y_{0}>0$ gives a de-Sitter expansion while for $Y_{0}<0$ eventually yields a static Universe. Of course, we should point out that the result (\ref{ae}) alone is not enough to determine the cosmological evolution even when it provides an analytic solution for $a(t)$, as for instance in (\ref{td}). This first integral (\ref{ae}) is then to be used along with the first Friedmann equation and the conservation laws of the given Theory in order to obtain a complete Cosmological scenario.

	\section{Conclusions}
	
	We have fully classified the various cases of Perfect Cosmological Hyperfluids, discussing also their conservation laws and subsequently some of their kinematical characteristics. In particular, we analysed the cases of pure dilation, pure shear and pure spin hyperfluids (along with some certain mixings) sourcing a Non-Riemannian Cosmological background. An alternative form of the conservation laws of Hyperfluids have also been given that explicitly contains the evolution of the dilation of shear, dilation and (implicitly) spin.
 As a specific example we have considered the case of a scalar field coupled to the connection for which we computed the dilation, shear and spin currents and have also computed explicitly the hyperfluid "equations of state", that is the relations among the hypermomentum variables of the Perfect Hyperfluid.
 In addition, special cases of some standard  Geometries sourced by such hyperfluids have been studied, such as Weyl, Weyl-Cartan etc. We have also given the definition of an incommpressible hyperfluid and discussed some of its properties.  

 On the second part of the paper, starting from the generalized acceleration equation (aka 2nd Friedmann or Raychaudhury equation), we have been able to extract a first integral of motion, evolving the scale factor and the kinematical features of the hyperfluid, namely the density, pressure and its microproperties (i.e. Hypermomentum variables). As an application we considered a Ricci flat geometry in the presence of a constant non-Riemannian field and showed how in this case this first integral provides an analytic solution for the scale factor.

 Related to the kinematical characteristics of the Perfect Cosmological Hyperfluid, there are two immediate conclusions worth stressing out. The first one is that a pure spin Hyperfluid can exist only if the fluid is not hypermomentum preserving. In other words, a pure spin hyperfluid supports a canonical energy-momentum tensor that is  necessary different from the metrical (i.e. Hilbert) energy-momentum tensor.  This conclusion is in perfect agreement with the fact that for particles carrying spin, the canonical and metrical energy-momentum tensors are different. The second kinematical fact (i.e. independent of the Gravitational action one considers) is that conservation conservation laws, namely the continuity equation and the evolution equations for the hypermomentum variables completely decouple from one another, for the case of a hypermomentum preserving, pure dilation fluid. More precisely, the modified continuity equation acquires its usual perfect fluid form  and the hypermomentum variable $\omega$ (or $\phi$), or better stated, the  
 dilation degree of freedom, evolves in a similar manner obeying the "dust-like" conservation law (\ref{dustlike}a).

 From the simple example of section $V$ we can draw two immediate conclusions for a scalar field coupled to the connection. The first is that the spin part of hypermomentum has no association to non-metricity, which is something to be expected. The second result, however, is not so obvious and shows us that torsion can be sourced from all parts of hypermomentum including also dilation a shear, the parts that are commonly attributed only to non-metricity. In addition, the same matter sector provides a concrete example on how the equations of state among the hypermomentum variables of the Hyperfluid can be obtained. Finally, we obtained yet another kinematical fact for non-Riemannian Cosmologies, in particular a first integral of motion, among the scale factor and the non-Riemannian degrees of freedom as given by eq. (\ref{ae}). The collection of our results then formalises the kinematical analysis of non-Riemannian effects in Cosmology and will be used to get a better grasp on the role of torsion and non-metricity (and their corresponding hypermomentum sources) in homogeneous and isotropic Cosmologies.

\section{Acknowledgements}

 This work was supported by the Estonian Research Council grant (SJD14).
		
	\bibliographystyle{unsrt}
	\bibliography{ref}

		\end{document}